# A Load-Balanced Call Admission Controller for IMS Cloud Computing

Ahmadreza Montazerolghaem[1], Mohammad Hossein Yaghmaee[1,2], Alberto Leon-Garcia[2], Mahmoud Naghibzadeh[1], Farzad Tashtarian[3]

[1]Department of Computer Engineering, Ferdowsi University of Mashhad, Mashhad, Iran
[2]Department of Electrical and Computer Engineering, University of Toronto, Toronto, Ontario, Canada
[3]Department of Information Technology, Mashhad Branch, Azad University, Mashhad, Iran
Ahmadreza.montazerolghaem@stu.um.ac.ir, hyaghmae@um.ac.ir, alberto.leongarcia@utoronto.ca, naghibzadeh@um.ac.ir,
f.tashtarian@mshdiau.ac.ir

*Abstract*— Network functions virtualization provides opportunities to design, deploy, and manage networking services. It utilizes Cloud computing virtualization services that run on high-volume servers, switches and storage hardware to virtualize network functions. Virtualization techniques can be used in IP Multimedia Subsystem (IMS) cloud computing to develop different networking functions (e.g. load balancing and call admission control). IMS network signaling happens through Session Initiation Protocol (SIP). An open issue is the control of overload that occurs when an SIP server lacks sufficient CPU and memory resources to process all messages. This paper proposes a virtual load balanced call admission controller (VLB-CAC) for the cloud-hosted SIP servers. VLB-CAC determines the optimal "call admission rates" and "signaling paths" for admitted calls along with the optimal allocation of CPU and memory resources of the SIP servers. This optimal solution is derived through a new linear programming model. This model requires some critical information of SIP servers as input. Further, VLB-CAC is equipped with an autoscaler to overcome resource limitations. The proposed scheme is implemented in SAVI (Smart Applications on Virtual Infrastructure) which serves as a virtual testbed. An assessment of the numerical and experimental results demonstrates the efficiency of the proposed work.

I. INTRODUCTION

IP Multimedia Subsystem (IMS) has been introduced by 3rd Generation Partnership Project (3GPP) standards to enable real-time communication services such as Voice over IP (VoIP), video communication, IP-based messaging and other new innovative multimedia communication services, on a full IP-based core network over any access technology.

IMS Cloud computing has been introduced to support high-quality multimedia applications in the cloud computing. Network Function Virtualization (NFV) and Cloud computing could be used to improve the scalability and elasticity of IMS deployments [1]. The work given in [2] introduces a set of three software architectures for efficient virtualization of IMS in different operating environments. A management architecture is used to simplify the deployment and runtime orchestration of such a virtual service on top of a Cloud infrastructure.

Call Session Control Function (CSCF) is the main part of IMS which is used to setup, maintain and release call sessions. The main protocol used for managing media sessions in CSCF is Session Initiation Protocol (SIP). SIP is an application layer signaling protocol for initiating, modifying, and tearing down multimedia sessions [3]. SIP uses UDP or TCP packets for different operations such as device registration and call setup/teardown. Call setup message or "INVITE" (briefly referred to as Call Request) is the most important request message, given that its transaction has a maximum process load on CPU server [4, 5]. Once a session is created, media are exchanged without passing through the servers. Hence, only the signaling messages would impose a load on the server. The SIP overload occurs when SIP server lacks enough resources such as CPU and memory [8], to process all messages [6, 7]. Considering the increasing application of this protocol, SIP serves should be equipped with overload control.

Cloud-hosted SIP servers offer flexibility, affordability, and reliability for communication requirements. Unlike traditional SIP servers, cloud-hosted SIP servers use the benefits of Cloud computing and eliminate the expensive setup and the large capital investment on hardware. Cloud-hosted SIP servers can use the dynamic resource provisioning and autoscaling features of the cloud computing to provide scalability, mobility, and reliability for the customers. Each SIP server can be implemented in a Virtual Machine (VM) in the cloud environment. As each VM has a predefined resource, the total capacity of SIP servers decreases when an overload situation occurs. This is because most of its processing resources are being used for rejecting or processing the messages that would be ultimately rejected [9]. SIP servers provide reliability by retransmission of the messages with an unconfirmed delivery [9] when functioning on unreliable transmission protocols such as UDP [8]. In this regard, a large set of retransmission timers are employed [6, 8]. Although this mechanism is useful in case of unreliable links, under overload conditions, it imposes high loads on the server and decreases efficiency [8]. This is because the redundant retransmissions and manipulation of the mentioned timers increase CPU and memory occupation and worsen server overload. On the other hand, the base mechanism of SIP protocol which lacks the required efficiency for overload control would be activated [6]. In this case, the server rejects new call request messages by issuing "503 service unavailable" message once reaching the maximum capacity (whose cost is comparable to providing a service). In addition, for servers configured statefully (as the dominant configuration), some state information is stored for each transaction. Having no supervision over the number of established calls, a server might engage its entire memory which degrades the performance. Therefore, to obtain maximum capacity and prevent overload occurrence, it is necessary to make optimum use of resources and prevent its waste.

The aim of this paper is to propose a method for optimal allocation of SIP servers' resources to the admitted calls. To this aim, we extend our recent work [10] by utilizing the benefits of NFV and the cloud computing to propose an optimized VLB-CAC. VLB-CAC runs on a virtual machine on the cloud while communicating with all SIP servers. Meanwhile, it collects statistics about the remaining resources of SIP servers. VLB-CAC is responsible to find the optimal call acceptance rate for each SIP server by solving an optimization problem which prevents overload. VLB-CAC uses the autoscaling capability of cloud computing to overcome resource limitations, using the cloud capacities in resources virtualization. The main contributions of this work are summarized as follows:

(1) We develop an optimized virtual load balancer and admission controller for SIP servers by utilizing the virtualization services in cloud computing;

(2) We prove that the overload control problem for $n$ servers with limited resource is NP-hard;

(3) We propose a mathematical model which optimizes resource usage and maximizes throughput;

(4) We propose a novel autoscaling scheme at VM level, which sufficiently scales the resources, while attempting to maximize call admission rate;

(5) We implemented the proposed work on a real two-tier cloud computing [11].

The rest of the paper is organized as follows: in Section II, we provide an overview of the related works. System model and problem formulations is given in Section III. Section IV presents the proposed method. The performance evaluation, conclusion and future work are given Section VI.

## II. RELATED WORK

There are different approaches for improving SIP performance under overload conditions. These approaches are classified as below:

- Load balancing: Distribution of new input traffic over the SIP servers based on their accessible capacity using a "Load Balancer" [4, 12, and 13] is called load balancing. This approach reduces the probability of overload. As the whole signaling traffic of SIP network passes through load balancer, performance is reduced. As a result, the load balancer itself is threatened by the risk of overload. Therefore, its performance needs to be enhanced by the available techniques.

- Overload control: These methods are divided into local and distributed methods. In local methods, the overloaded server has an independent control over its resource usage without the need for interaction with other network servers. The criteria for identifying the overload in some of these methods are queue length and CPU usage level. According to these criteria, a set of thresholds is defined, exceeding which makes the server enter overload stage so that it starts to reject the incoming calls. The main drawback of this method is that the cost of call rejection cannot be ignored, and when dealing with heavy overloads, the server must use its resources for rejecting the excess calls [5]. On the other hand, depending on whether upstream servers detect overload or informed via the overloaded downstream servers, distributed methods are categorized into explicit and implicit methods [7]. Explicit methods are classified into rate-based, loss-based, signal-based, window-based, and on/off control techniques [14]. Within the rate-based techniques, the downstream server controls the delivery rate from upstream servers [9, 15]. In the loss-based technique, downstream server frequently measures its current load and accordingly requests the upstream servers to reduce their transmitted load [16]. In window-based methods, unless there are empty slots in upstream server window, the load is not transmitted to the downstream server. The main issue of the later methods is window size adjustment, which can be achieved using the feedback from the downstream server [3, 6, and 17]. In signal-based methods, the upstream server reduces its transmission rate when receiving "503 Service Unavailable" message in order to prevent further transmission of 503 messages from the downstream server [18]. Unlike signal-based method which do not employ "Retry-After" header, a given server can either hold off or on its received load within the on/off control method by transmission of Retry-After feedback [19]. As opposed to explicit methods, the absence of responses or the loss of packets is used to detect overload in implicit methods.

- Retransmission rate variation: These methods review retransmission mechanism of SIP by studying servers' buffer size [8, 20]. By limiting the dedicated memory of the server, admitting over-capacity calls can be avoided. This policy loses efficiency once the call rate rises, as the server processor is forced to analyze the messages to recognize their content. Therefore, the server saturates, (typically) under higher loads.

- Exploiting TCP flow control: These methods exploit the mechanism of preventing congestion for the purpose of overload control [21]. The setback of such methods is scalability and high delay. Further, congestion in TCP occurs due to limited bandwidth, whereas overload in SIP happens due to limited processing capacity of servers' CPU [8].

As mentioned earlier, cloud-hosted SIP servers use the benefits of cloud computing. Moreover, they eliminate the expensive setup and large capital investment on hardware. Therefore, SIP servers can be implemented as VMs in the cloud. Resource management and call admission control are important issues in cloud computing. In [22], a distributed hierarchical framework based on a mixed-integer nonlinear optimization of resource management across multiple timescales is proposed. The main goal is to set resource allocation policies for virtualized cloud environments that satisfy performance and availability guarantees and minimize energy costs in very large cloud service centers. In [23] a centralized hierarchical cloud-based multimedia system (CMS) has been considered. The CMS consists of a resource manager, cluster heads, and server clusters in which each server cluster only handles a specific type of multimedia task, and each client requests a different type of multimedia service at a different time. The problem has been modeled as an integer linear programming and has been solved by an efficient genetic algorithm. The authors in [24] formulate and discuss load balancing problem addressing VoIP in the cloud computing federation and propose a new distributed adaptive power aware load balancing algorithm for VoIP cloud.

Based on the above-mentioned points, the disadvantages of the present overload control approaches are: First, reliance merely on the local call rejection reduces throughput. Second, for the majority of explicit feedback-based methods, the continuous revision of the status and feedback calculation imposes complexity and overhead. Third is the delay in feedback arrival to upstream, which results in instability. Nonetheless, the precision of overload detection is higher compared to implicit methods. Therefore, overload detection, feedback generation, and running the overload control algorithm incurs CPU and memory usage costs and affects server throughput. To the best of our knowledge, autoscaling capability of the cloud computing has not been employed to scale up or down the VMs in case of sudden changes in offered load in IMS cloud computing.

## III. SYSTEM MODEL AND PROBLEM FORMULATION

The architecture of the SIP protocol consists of User Agent and SIP Server. A User Agent Client creates a SIP request and sends it to a User Agent Server. The request traverses through one or more SIP servers in a SIP network. The main purpose of a SIP server is to route each request to its destination [9]. Since the requests are routed hub by hub, and generally there are several servers from source to destination (which implies various path between the nodes), factoring in the routing request from the SIPs would be beneficial to lowering resource for servers and avoiding loops. The response traces back the path that the request has been taken. An "INVITE" request together with a "BYE" request initialize and terminate the call, respectively. As SIP is call-oriented, it can only reject the "INVITE" requests by the SIP server if unwilling (or unable) to forward the requests. The user is connected to the server located closest to it. The first server to which the call arrives, decides about the admission of the call. Moreover, the middle servers route the call to the last server (or the target server).

Consider a two-tier cloud consisting of one core that represents a cloud with large number of computing resources, typically a large data center and some smart edges that represent a cloud with a limited set of computing resources, typically a small data center. The smart edges are geographically close to end users thus resulting in low network latency. To facilitate the movement of computation tasks, the core and smart edge(s) are connected by a fast network. A virtualized network function, or VNF, may consist of one or more VMs running different software and processes, on top of standard high-volume servers, switches and storage, or cloud computing infrastructure.

As shown in Fig. 1, the system in general consists of a set of $n$ SIP servers located at different VMs with limited processing and memory resources. The VMs are employed instead of having custom hardware appliances for each network function. At each smart edge, there are one or more SIP servers implemented in a VM. At the core of the cloud, there is a Virtual Load Balanced Call Admission Controller (VLB-CAC), which acts as the central controller, collects information about the available resources and the offered load of all SIP servers, and through running an optimization problem, finds the optimal resource allocation and broadcasts the results to all SIP servers. Each server uses its resources to initiate a session between the local users and users in other domains. In this paper, it is assumed that the binary symmetric matrix $L_{ij}, i,j \in \{1,...,n\}$ represents the topology of the SIP server network. In this matrix, $L_{ij} = 1$ implies the presence of a communication link via SIP trunk between servers $i$ and $j$, while $L_{ij} = 0$ indicates that the two given SIP servers are not adjacent through a direct SIP trunk. Note that the main diagonal elements of this matrix are all zeros. SIP trunk is a service for making a connection between the VoIP network devices that work with SIP protocol. In fact, a SIP trunk is a virtual connection over the public internet or a Virtual Private Network (VPN) that connects VoIP equipment. Assume a two-dimensional array $\mathbb{C}$ with the size of $(n \times n)$, where $\mathbb{C}^{ii}$ denotes the number of local calls in server $i$ (where both caller and callee are registered on the same server) and $\mathbb{C}^{ij}$ shows the number of outbound calls from server $i$ to server $j$. Let $C^{ij}$ be the optimal number of admitted calls established from server $i$ to server $j$ where $C^{ij} \leq \mathbb{C}^{ij}$. Regarding the optimal value of $C^{ij}$, $R_{kl}^{ij}$ shows the number of calls from origin $i$ to destination $j$, which must be relayed from server $k$ to server $l$ (see Fig. 2).

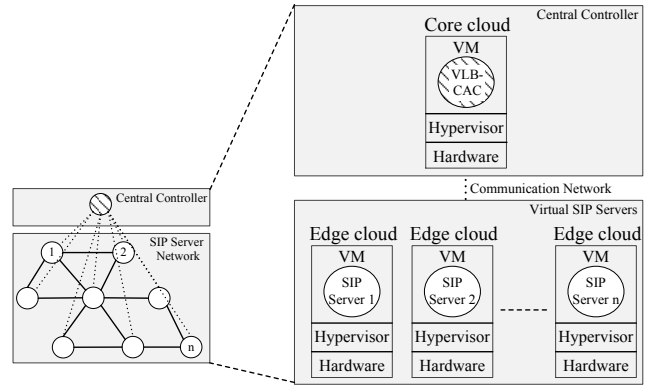

Fig. 1. A simple model of the proposed system in a two-tire cloud

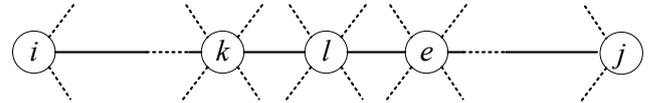

Fig. 2. Transmitting calls from the origin $i$ to the destination $j$ assisted by SIP servers $k$ and $l$

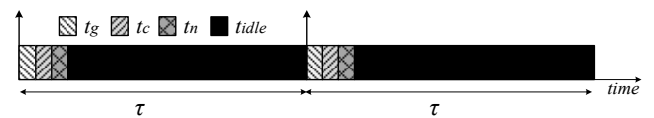

Fig. 3. The duty cycle of VLB-CAC

To perform its operation, each server $i$ relies on its remaining CPU and memory resources, which are denoted as $P_i$ and $M_i$, respectively. The duty cycle of VLB-CAC is depicted in Fig. 3. At the beginning of each time slot $\tau$, all servers transmit how much the resources are remaining and the number of new local and outbound requests to the VLB-CAC through UDP packets. This critical information is gathered in every $t_g$ units of time. By using the linear program (LP) model proposed in the next part of this paper, VLB-CAC determines the optimal values of $C^{ij}$ and $R_{kl}^{ij}$ in $t_c$ and then broadcasts the results to the servers in $t_n$ (Fig. 3). After that, VLB-CAC enters into the idle state and waits $t_{idle}$ units of time for the next time slot $\tau$. In fact, a new incoming call during $\tau$ should be entered into the hold-on state by the SIP server until the next $\tau$. Compared to the duration of a call, this time, is negligible which can also be shortened by reducing $\tau$.

## IV. VLB-CAC

In this section, we propose to design the VLB-CAC controller in a way to optimally determine $C^{ij}$ and $R_{kl}^{ij}$, based on a new linear programming model. Prior to proposing the model, we prove that the problem of overload controlling in SIP networks is in the form of a mixed integer linear program and is, therefore, NP-hard.

*Theorem 1*: Overload control problem for a set of $n > 2$ SIP servers with limited CPU and memory resources is in the form of a mixed integer nonlinear program (MINLP).

*Proof*: Let $B_{kl}^{ij}$ be a binary variable. Assume that $B_{kl}^{ij} = 1$ denotes that a call request from server $i$ to server $j$ could be transmitted from servers $k$ to server $l$ and $B_{kl}^{ij} = 0$ denotes that server $k$ and $l$ do not participate in transferring request from server $i$ to server $j$. Given the limited resources in the servers, the optimal values of $B_{kl}^{ij}$, $C^{ij}$, and $R_{kl}^{ij}$ could be obtained through the following MINLP model:

$$\max \sum_{i=1}^{n}\sum_{j=1}^{n} C^{ij} \qquad (1)$$

Subject to:

$$C^{ij} \leq \mathbb{C}^{ij}, \qquad \forall i,j \text{ (I)}$$
$$\sum_{k=1}^{n} B_{kl}^{ij} R_{kl}^{ij} = \sum_{e=1}^{n} B_{le}^{ij} R_{le}^{ij}, \qquad \forall i,j,l, i \neq l, j \neq l \text{ (II)}$$
$$\sum_{k=1}^{n} B_{kl}^{ij} R_{kl}^{il} = C^{il}, \qquad \forall i,j,l, i \neq l \text{ (III)}$$
$$\sum_{e=1}^{n} B_{le}^{ij} R_{le}^{lj} = C^{lj}, \qquad \forall i,l,j, j \neq l \text{ (IV)}$$
$$R_{kl}^{ii} = 0, \qquad \forall i,k,l \text{ (V)}$$
$$R_{ki}^{ij} = 0, \qquad \forall i,j,k \text{ (VI)}$$
$$B_{kl}^{ij} - L_{kl} \leq 0, \qquad \forall i,j,k \text{ (VII)}$$
$$\alpha_1 C^{ll} + \alpha_2 (\sum_{i=1}^{n}\sum_{j=1}^{n}\sum_{k=1}^{n}(B_{lk}^{ij} R_{lk}^{ij} + B_{kl}^{ij} R_{kl}^{ij})) \leq P_l, \qquad \forall l \text{ (VIII)}$$
$$\beta_1 C^{ll} + \beta_2 (\sum_{i=1}^{n}\sum_{j=1}^{n}\sum_{k=1}^{n}(B_{lk}^{ij} R_{lk}^{ij} + B_{kl}^{ij} R_{kl}^{ij})) \leq M_l, \qquad \forall l \text{ (IX)}$$

Variables: $B_{kl}^{ij} \in \{0,1\}, C^{ij}, R_{kl}^{ij}, P_l, M_l \geq 0, \forall k,l,i,j$.

Note that Constraint (I) limits the number of admitted calls to the number of existing call requests. Constraint (II) offers a trade-off between the input and output flows; meaning that it enforces equal total inward and outward flows between each pair of origin and destination for server $l$. Constraint (III) aggregates the total input flows to server $l$ from origin $i$ passing through its neighbors. The next constraint distributes the total output flows from server $l$ to server $j$ among its neighbors. Constraint (V) prevents a flow with the same origin and destination. Constraint (VI) prevents the creation of certain loops in a given path, which will be discussed in Section IV. A. Constraint (VII) limits the binary variable $B_{kl}^{ij}$ in such a way that $B_{kl}^{ij}$ is zero for $L_{kl} = 0$, and it could be either zero or one if $L_{kl} = 1$. The next two constraints take into account the limited SIP resources of servers. Constraint (VIII) allocates the residual processing power of server $l$ to establish local and outbound calls with coefficients $\alpha_1$ and $\alpha_2$. Similarly, Constraint (IX) allocates the residual memory of server $l$ with coefficients $\beta_1$ and $\beta_2$. It is noteworthy that the two variables $R_{lk}^{ij}$ and $R_{kl}^{ij}$ contribute evenly to the servers' resources. To estimate parameters $\alpha_1$, $\alpha_2$, $\beta_1$, and $\beta_2$, two linear programming models are proposed in Section IV.C.

Although the objective function is linear, the nonlinear constraints and the binary variables $B_{kl}^{ij}$ render the model an MINLP which is generally NP-hard and unsolvable in polynomial time [25, 26]. To overcome this drawback, we propose certain modifications, which remove the nonlinearity and allow the problem to take the form of an LP.

*A. Proposed Heuristic Method*

Before discussing the proposed method, we should elaborate on loop characterization. We divide the loops between the origin and the destination into two types, namely k-hop source loop (SL) and k-hop non-source loop (nSL). In the former type, the source server participates in the loop, whereas, in the later, the loop is created by servers other than the source server. The proposed model in (1) can only bind SLs as it applies Constraints (V) and (VI). Fig. 4 provides a visualization of the two loop types.

Although the proposed model in (1) increases the total number of admitted calls, it can be easily shown that if $\sum_{i=1}^{n}\sum_{j=1}^{n} C^{ij}$ is considered as the objective function, it is not possible to prevent nSLs. Given also that the nonlinear model is NP-hard, the objective function should be modified in such a way that not only maximizes the total number of admitted calls but also minimizes CPU and memory usages (Multi-objective). In multi-objective problems, several objective functions are optimized simultaneously.

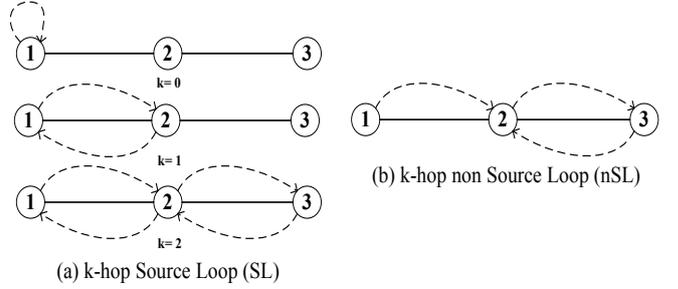

Fig. 4. SL and nSL between the origin and destination flow path. Note that '1' and '3' are the origin and destination servers, respectively.

An advantage of such a modification is that it would prevent nSLs since loops creation requires resources. To this end, we consider the following LP model:

$$\max \gamma \frac{\sum_{i=1}^{n}\sum_{j=1}^{n} C^{ij}}{\sum_{i=1}^{n}\sum_{j=1}^{n} \mathbb{C}^{ij}} - \varphi (\frac{\sum_{l=1}^{n} p_l}{\sum_{l=1}^{n} P_l} + \frac{\sum_{l=1}^{n} m_l}{\sum_{l=1}^{n} M_l}) \qquad (2)$$

Subject to:

$$C^{ij} \leq \mathbb{C}^{ij}, \qquad \forall i,j \text{ (I)}$$
$$\sum_{k=1}^{n} L_{kl} R_{kl}^{ij} = \sum_{e=1}^{n} L_{le} R_{le}^{ij}, \qquad \forall i,j,l, i \neq l, j \neq l \text{ (II)}$$
$$\sum_{k=1}^{n} L_{kl} R_{kl}^{il} = C^{il}, \qquad \forall i,l, i \neq l \text{ (III)}$$
$$\sum_{e=1}^{n} L_{le} R_{le}^{lj} = C^{lj}, \qquad \forall l,j, j \neq l \text{ (IV)}$$
$$R_{kl}^{ii} = 0, \qquad \forall i,k,l \text{ (V)}$$
$$R_{ki}^{ij} = 0, \qquad \forall i,j,k \text{ (VI)}$$
$$\alpha_1 C^{ll} + \alpha_2 (\sum_{i=1}^{n}\sum_{j=1}^{n}\sum_{k=1}^{n}(L_{lk} R_{lk}^{ij} + L_{kl} R_{kl}^{ij})) \leq p_l, \qquad \forall l \text{ (VII)}$$
$$\beta_1 C^{ll} + \beta_2 (\sum_{i=1}^{n}\sum_{j=1}^{n}\sum_{k=1}^{n}(L_{lk} R_{lk}^{ij} + L_{kl} R_{kl}^{ij})) \leq m_l, \qquad \forall l \text{ (VIII)}$$
$$p_l \leq P_l, \qquad \forall l \text{ (IX)}$$
$$m_l \leq M_l, \qquad \forall l \text{ (X)}$$

Variables: $C^{ij}, R_{kl}^{ij}, p_l, m_l \geq 0, \forall i,j,k,l$.

Note that the connectivity between two servers is developed using the matrix $L_{ij}$. A loose upper bound for the optimal CPU and memory usages are given by $p_l$ and $m_l$, respectively, and based on Constraints (IX) and (X), they could reach the maximum $P_l$ and $M_l$ (the strict power bound). Constraints (VII) and (VIII) would further help the objective function prevent the creation of extra $R_{kl}^{ij}$, SL and nSL loops. Coefficients $\gamma$ and $\varphi$ indicate the significance of variables $C^{ij}$ and resource usage ($p_l$ and $m_l$) in the objective function, respectively. In other words, they offer a trade-off between the resource usage and network throughput. Ultimately, this model enhances the total throughput of the servers without any overload by using the optimal resource allocation. Here, the overall throughput could be considered as the total flow passing through the servers. Therefore, taking into account the servers' resources, VLB-CAC tries to maximize call admission through distributing calls between the servers.

*B. VLB-CAC with Proactive Autoscaling*

In model (2), the enhancement of the call admission rate is dependent on server resources and the number of call requests. To overcome the resource limitations, we propose a novel resource autoscaling scheme at the VM level (Fig. 5). It attempts to scale just enough resources to minimize resource waste, while tries to maximize call admission rate regarding the number of call requests. Autoscaling is a technique for dynamically adjusting the resources of VMs to an appropriate *flavor* in response to an increase in demand. A flavor is an available resource configuration for a VM on a server. It defines the size of a virtual server that can be launched (see Table III).

Fig. 5 shows that this scheme includes four modules. Since autoscaling lasts for a few seconds [27], the proposed method must be predictive-driven. Predictive autoscaling algorithms predict future system behaviour and adjust resources in advance to meet the future needs. Therefore, Module 1 predicts the number of the call requests for the next $\tau$ ($\widehat{\mathbb{C}^{ij}}(n_\tau + 1)$). Module 2 specifies the required resources for admission of the requests in each server (($p_l^*, m_l^*)_{(n_\tau+1)}$). Module 3 selects the most appropriate flavor for each VM. The selected flavor might be more than the resources specified for the VM. Considering the selected flavors, Module 4 specifies an upper bound for the admitted calls in each server which has been obtained by Module 2. In this way, the method will be less sensitive to the predicted results. Eventually, the scheme makes a resource scaling decision for each server as a rule that is up or down or NOP (no operation) in each $\tau$.

*1) Module 1 - NLMS prediction system*

We focus on proactive autoscaling that is based on predictions of future workload ($\widehat{\mathbb{C}^{ij}}(n_\tau + 1)$) based on the past workload ($\underline{\mathbb{C}}^{ij}(n_\tau)$). The most important method in this area is time-series analysis. Time-series analysis could be used to detect repeating patterns in the workload or to estimate future values for resource allocation. Consequently, the scaling action is done in advance. In order to apply time-series analysis in autoscaling domain, a certain performance metric will be periodically sampled at fixed intervals ($\mathbb{C}^{ij}(n_\tau)$). The result will be a time-series containing a sequence of last observations. Time-series methods extrapolate this sequence to predict future values. Some of the methods used for this purpose are Moving Average, Autoregression, Autoregressive–Moving-Average (ARMA), exponential smoothing, and machine learning techniques [27].

We utilize the work of [28] that proposes a comparative analysis among prediction techniques, used for developing a dynamic prediction based the resource allocation strategy. Among these techniques, the Normalized Least Mean Square (NLMS) predictor is the one providing the best trade-off between complexity, accuracy, and responsiveness.

The general problem of prediction can be stated as follows: given a set of observations of a stochastic process $x(n)$, generate an estimation $\hat{x}(n + k)$ of the value $x(n + k)$ that the process $x$ will assume $k$ steps ahead. Given a vector of $p$ observations, $\underline{x} = [x(n), x(n - 1), \dots x(n - p + 1)]$, the predicted value $\hat{x}$ is obtained by $\hat{x} = \psi(\underline{x})$ where the function $\psi$ is called the *predictor*.

Various categories of predictors have been studied; however, considering the constraint on the complexity, the linear prediction category is the best suited for our aim. A linear prediction happens whenever the function $\psi(\underline{x})$ is linear. Putting it differently, the problem is to specify the impulse response $h(n)$ of the linear filter $h$ such that:

$$\hat{x}(n + k) = x(n) \otimes h(n) = \sum_{i=0}^{p-1} h(i)x(n - i) \quad (3)$$

The filters coefficients can be determined according to arbitrary optimality criteria. One of the widely adopted prediction technique is the so called Linear Minimum Mean Square Error (LMMSE) predictor, in which the values $h(n)$ are derived by minimizing the Mean Square Error of prediction:

$$\mathbb{E}[\epsilon^2(n)] = \mathbb{E}[(x(n + k) - \hat{x}(n + k))^2] \quad (4)$$

The problem of this predictor is that the derivation of the LMMSE filter needs the knowledge of at least $p$ values of the autocorrelation function of the stochastic process $x(n)$ and the inversion of a $p \times p$ matrix. These facts make LMMSE inappropriate for being used as on-line predictive method. Therefore, we consider the NLMS method, which is based on an adaptive mechanism. It does not require previous knowledge of the autocorrelation structure of the stochastic sequence [29]. The algorithm scheme is shown in Fig. 6.

The filter coefficients are time varying and are tuned on the basis of the feedback information carried by the error $\mathcal{E}(n)$. We define the vector of filter coefficients at time $n$ with $\underline{h}_n$. The values of $\underline{h}$ adapt dynamically in order to decrease the Mean Square Error. Note that $\mathcal{E}(n) = x(n + k) - \hat{x}(n + k)$ and $\underline{x}_n = [x(n), x(n - 1), \dots x(n - p + 1)]$.

The NLMS works as follows: Initialize the coefficient $\underline{h}_0$; and for each new data, update the filter $h(n)$ according to the recursive equation.

$$\underline{h}_{n+1} = \underline{h}_n + \mu \frac{\mathcal{E}(n)\underline{x}_n}{\|\underline{x}_n\|^2} \quad (5)$$

where $\|\underline{x}_n\|^2 = \underline{x}_n \underline{x}_n^T$ and $\mu$ is a fixed parameter called step size. Pursuant to [29], NLMS converges in the mean to LMMSE predictor as long as $0 < \mu < 2$. At time $n$, the values $x(n + k)$, hence $\mathcal{E}(n)$, are not known. So, the value $\mathcal{E}(n - k)$ is used, instead, and the one step NLMS predictor update equation becomes:

$$\underline{h}_{n+1} = \underline{h}_n + \mu \frac{\mathcal{E}(n - 1)\underline{x}_{n-1}}{\|\underline{x}_{n-1}\|^2} \quad (6)$$

For our problem, the NLMS algorithm formulation can be summarized as:

Parameters: $p$ = filter order, $\mu$ = step size

Initialization: $\underline{h}(0) = 0$

Computation: For $n_\tau$:

$$\underline{\mathbb{C}}^{ij}(n_\tau) = [\mathbb{C}^{ij}(n_\tau), \mathbb{C}^{ij}(n_\tau - 1), \dots \mathbb{C}^{ij}(n_\tau - p + 1)] \quad (7)$$

$$\widehat{\mathbb{C}^{ij}}(n_\tau + 1) = \underline{h}(n_\tau) \times \underline{\mathbb{C}}^{ij^T}(n_\tau) \quad (8)$$

$$\underline{h}(n_\tau) = \underline{h}(n_\tau - 1) + \mu \frac{\mathcal{E}(n_\tau - 1)\underline{\mathbb{C}}^{ij}(n_\tau - 1)}{\|\underline{\mathbb{C}}^{ij}(n_\tau - 1)\|^2} \quad (9)$$

$$\mathcal{E}(n_\tau) = \mathbb{C}^{ij}(n_\tau) - \widehat{\mathbb{C}^{ij}}(n_\tau) \quad (10)$$

$$\|\underline{\mathbb{C}}^{ij}(n_\tau)\|^2 = \underline{\mathbb{C}}^{ij}(n_\tau)\underline{\mathbb{C}}^{ij^T}(n_\tau) \quad (11)$$

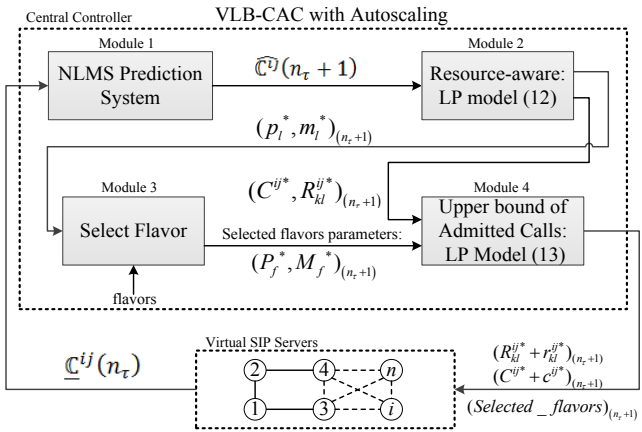

Fig. 5. VLB-CAC with autoscaling scheme

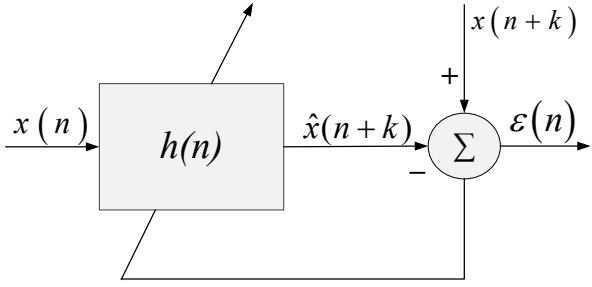

Fig. 6. NLMS algorithm

Where $\mathbb{C}^{ij}(n_\tau)$ and $\widehat{\mathbb{C}^{ij}}(n_\tau + 1)$ are input and output of Module 1, respectively. The NMLS predictor needs the configuration of two parameters: the order $p$ and the step size $\mu$. The order $p$ has been chosen to be 30, since it produced a good performance based on the results analysis. In the case of $\mu$, it is relevant to note that one of the important advantages of using NLMS is that it is less sensitive to the step size compared with other linear predictors. In our experiments, we chose the step value of 0.8 as a trade-off between fast convergence and responsiveness to input change. We run various tests to tune this parameter. For small value of $\mu$, we apperceive that the prediction function slowly converges and is unable to follow sudden workload increase. For values of $\mu$ greater than 0.5, the prediction function is not sensitive to step size and results in a quicker response to workload changes.

*2) Module 2 – Resource aware*

The linear model presented below in (12) specifies the minimum amount of the resources required for the admission of the whole call requests in each server predicted for the next $\tau$ $((p_l^*, m_l^*)_{(n_\tau+1)})$. It also specifies the optimal values of $R_{kl}^{ij}$ and $C^{ij}$ for the next $\tau$ $((C^{ij^*}, R_{kl}^{ij^*})_{(n_\tau+1)})$. Considering Constraint (I), the value of $(C^{ij^*})_{(n_\tau+1)}$ is equal to the number of the predicted requests by Module 1. Notice that this model is adopted from model proposed in (2).

*LP model: Resource-aware*
$$\min \ \frac{\sum_{l=1}^n p_l}{\sum_{l=1}^n P_l} + \frac{\sum_{l=1}^n m_l}{\sum_{l=1}^n M_l} \tag{12}$$
Subject to:
$C^{ij} = \widehat{\mathbb{C}}^{ij}$, $\forall i,j$ (I)
Constraints II – VIII of model in (2), (II - VIII)
Variables: $C^{ij}, R_{kl}^{ij}, p_l, m_l \geq 0$, $\forall i,j,k,l$.

*3) Module 3 - Select flavor*

Since creating a new flavor appropriate for the output of Module 2 takes more time than resizing the existing flavors, we assume that there are $Q$ flavors with specific resources in the form of $flavor[Q] = [P_f, M_f]$. Module 3 selects the most appropriate flavor for the VMs in each server (Algorithm 1). Each VM resizes its flavor before reaching the next $\tau$.

*4) Module 4 - Upper bound of admitted calls*

As there is a gap between each selected flavor and the resources for the VMs in each server, an upper bound is determined for admission of the call requests by the model shown in (13), which is a derivative of the model in (2). $(C^{ij} + c^{ij})$ and $(R_{kl}^{ij} + r_{kl}^{ij})$ are the upper bounds for admission of the call requests based on the selected flavors. In this way, the possibility of hold-on for the new calls reduces.

**Algorithm 1** Select Appropriate Flavors for next $\tau$
1: **Input:**
2: $flavor[Q] = [P_f, M_f]$
3: $p_l^*, m_l^*$
4: **Output:**
5: **for** $j = 1$ **to** number of SIP servers (n)
6:   **for** $i = 1$ **to** number of flavors ($Q$)
7:     **if** $flavor[i].[P_f] \geq p_j^*$ **then**
8:       **if** $flavor[i].[M_f] \geq m_j^*$ **then**
9:         $Selected\_flavor[j] = flavor[i]$
10:         break
11:       **end if**
12:     **end if**
13:   **end for**
14: **end for**

*LP model: Upper bound of admitted calls*
$$\max \ \sum_{i=1}^n \sum_{j=1}^n c^{ij} \tag{13}$$
Subject to:
$\sum_{k=1}^n L_{kl}(R_{kl}^{ij^*} + r_{kl}^{ij}) =$    $\forall i,j,l, i \neq l, j \neq l$ (I)
$\sum_{e=1}^n L_{le}(R_{le}^{ij^*} + r_{le}^{ij})$ ,
$\sum_{k=1}^n L_{kl}(R_{kl}^{il^*} + r_{kl}^{il}) = (C^{il^*} + c^{il})$,   $\forall i,l, i \neq l$ (II)
$\sum_{e=1}^n L_{le}(R_{le}^{lj^*} + r_{le}^{lj}) = (C^{lj^*} + c^{lj})$,   $\forall l,j, j \neq l$ (III)
$R_{kl}^{ii^*} + r_{kl}^{ii} = 0$,   $\forall i,k,l$ (IV)
$R_{ki}^{ij^*} + r_{ki}^{ij} = 0$,   $\forall i,j,k$ (V)
$\alpha_1(C^{ll^*} + c^{ll}) + \alpha_2(\sum_{i=1}^n \sum_{j=1}^n \sum_{k=1}^n (L_{lk}(R_{lk}^{ij^*} + r_{lk}^{ij}) + L_{kl}(R_{kl}^{ij^*} + r_{kl}^{ij}))) \leq P_f^*$,   $\forall l$ (VI)
$\beta_1(C^{ll^*} + c^{ll}) + \beta_2(\sum_{i=1}^n \sum_{j=1}^n \sum_{k=1}^n (L_{lk}(R_{lk}^{ij^*} + r_{lk}^{ij}) + L_{kl}(R_{kl}^{ij^*} + r_{kl}^{ij}))) \leq M_f^*$,   $\forall l$ (VII)
Variables: $c^{ij}, r_{kl}^{ij} \geq 0$, $\forall i,j,k,l$.

VLB-CAC sends the values of $(C^{ij^*} + c^{ij^*})_{(n_\tau+1)}$ and $(R_{kl}^{ij^*} + r_{kl}^{ij^*})_{(n_\tau+1)}$ and the selected flavors for the next $\tau$ to the servers. Each server has an autoscaling rule table, which resizes its VM (Table I).

TABLE I.    AUTOSCALING RULE TABLE

| Rules | Actions |
|---|---|
| (selected flavor)$_{(n_\tau)}$ = (selected flavor)$_{(n_\tau+1)}$ | NOP |
| (selected flavor)$_{(n_\tau)}$ < (selected flavor)$_{(n_\tau+1)}$ | Scale UP |
| (selected flavor)$_{(n_\tau)}$ > (selected flavor)$_{(n_\tau+1)}$ | Scale Down |

*C. Calculation of α and β*

We propose the following two linear models to approximate the coefficients $\alpha$ and $\beta$. To this end, consider the quadruple $(\acute{C}_q^{ii}, \acute{R}_{ij,q}^{ij}, \acute{p}_q, \acute{m}_q)_{q=1,\ldots,h}$, where $\acute{C}_q^{ii}$ is the number of local calls in server $i$, $\acute{R}_{ij,q}^{ij}$ is the number of outbound calls of server $i$, $\acute{p}_q$ and $\acute{m}_q$ are CPU and memory usages in server $i$, respectively. We acquire all $h$ samples of this quadruple by conducting tests on a real testbed[1]. Consider the following:

$$\min \ \sum_{q=1}^n x_q \tag{14}$$
Subject to:
$\acute{p}_q - (\alpha_1 \acute{C}_q^{ii} + \alpha_2 \acute{R}_{ij,q}^{ij}) \leq x_q$,   $q = 1,\ldots,n$ (I)
$\alpha_1 + \alpha_2 = \frac{max\{\acute{p}_q\}}{max\{\acute{C}_q^{ii}\}}$,   $q = 1,\ldots,n$ (II)
Variables: $\alpha_1, \alpha_2, x_q \geq 0$, $q = 1,\ldots,n$,
$$\min \ \sum_{q=1}^n y_q \tag{15}$$
Subject to:
$\acute{m}_q - (\beta_1 \acute{C}_q^{ii} + \beta_2 \acute{R}_{ij,q}^{ij}) \leq y_q$,   $q = 1,\ldots,n$ (I)

---
[1] The tests are performed on SAVI testbed.

$$\beta_1 + \beta_2 = \frac{max\{\acute{m}_q\}}{max\{\acute{C}_q^{ii}\}}, \quad q = 1, \dots, n \quad (II)$$

Variables: $\beta_1, \beta_2, y_q \geq 0$, $q = 1, \dots, n$.

Considering the first constraints of (14) and (15), the objective functions are designed to minimize the sum of the difference between measured values of $\acute{p}_q$ and $\acute{m}_q$ with $\alpha_1 \acute{C}_q^{ii} + \alpha_2 \acute{R}_{ij,q}^{ij}$ and $\beta_1 \acute{C}_q^{ii} + \beta_2 \acute{R}_{ij,q}^{ij}$, respectively. In other words, $x_q$ denotes the difference between measured $\acute{p}_q$ and $\alpha_1 \acute{C}_q^{ii} + \alpha_2 \acute{R}_{ij,q}^{ij}$, and $y_q$ denotes the difference between $\acute{m}_q$ and $\beta_1 \acute{C}_q^{ii} + \beta_2 \acute{R}_{ij,q}^{ij}$, respectively. Objective functions (14) and (15) seek the optimal values for $\alpha$ and $\beta$ by minimizing $x_q$ and $y_q$. This is because to minimize to the objective functions, the first constraints has to diminish $\acute{p}_q - (\alpha_1 \acute{C}_q^{ii} + \alpha_2 \acute{R}_{ij,q}^{ij})$ and $\acute{m}_q - (\beta_1 \acute{C}_q^{ii} + \beta_2 \acute{R}_{ij,q}^{ij})$ in (15). Hence, $\alpha_1 \acute{C}_q^{ii} + \alpha_2 \acute{R}_{ij,q}^{ij}$ approaches $\acute{p}_q$ only when $\alpha$ is optimized. Note that the second constraints are imposed for the purpose of normalization of the first constraints.

## V. PERFORMANCE EVALUATION

### A. Simulations and Numerical Results

To simulate and assess the performance of the proposed model in Eq. (2), we implement our work in MATLAB. We considered the topology shown in Fig. 7; moreover, the three scenarios of Table II are investigated with different $\gamma$ and $\varphi$ (represented in four Cases: *f1* to *f4*). In all scenarios, calls are generated in random with the normal distribution ($\mathbb{C}^{ij}$).

The values of $M_l$ and $P_l$ ($l: 1, \dots, n$) are set to 100 for all servers. $\tau$ is set to 3 seconds for all runs. Furthermore, for $\alpha$ and $\beta$ coefficients, the obtained values in Section V.B.2 are considered. These values are 0.07841 and 0.02158 for $\alpha_1$ and $\alpha_2$ and 0.06998 and 0.01997 for $\beta_1$ and $\beta_2$, respectively.

Figs. 8 and 9 illustrate optimal values of $p_l$ and $m_l$ for all servers. In addition, Fig. 10 presents optimal call admission rates for different cases in different scenarios. In these figures, for all scenarios, by considering different cases (*f1* to *f4*), the resource usage and call admission rates show an increasing trend.

In this regard, the ratio between parameters $\gamma$ and $\varphi$ illustrates the importance of call admission or resource preservation. In case *f1*, resource preservation is more significant as compared to case *f4*. In contrast, in case *f4*, the maximum call admission is preferred, even if it results in higher resource consumption (Fig. 10). By making a trade-off between these parameters, it is possible to promote the call admission rates and optimal usage of the resources. Moreover, regarding the amount of load from Scenario 1 to Scenario 3, Figs. 8 and 9 reveal that resource usages would be raised; however, the input load of the network must be so high that even using the entire resource, it is not possible to respond to all input calls (Fig. 10, Scenario 3, cases *f3* and *f4*). In Scenario 1, the SIP servers do not enter the overload condition, as all input loads are admissible by using the limited amount of resources (Fig. 10, Scenario 1, cases *f3* and *f4*).

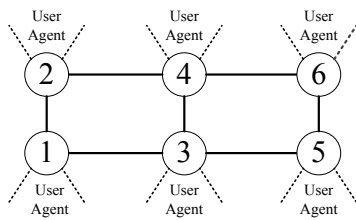

Fig. 7. The analyzed topology

TABLE II. THE SCENARIOS

| $\mathbb{C}^{ij}$ | | | | | | $\sum_{i=1}^{n}\sum_{j=1}^{n}\mathbb{C}^{ij}$ |
|---|---|---|---|---|---|---|
| 10 | 20 | 48 | 30 | 60 | 8 | |
| 10 | 50 | 20 | 54 | 48 | 54 | |
| 44 | 44 | 100 | 30 | 40 | 12 | 1300 |
| 10 | 20 | 30 | 46 | 50 | 14 | |
| 50 | 50 | 30 | 20 | 54 | 20 | |
| 50 | 40 | 25 | 40 | 54 | 15 | |
| Scenario 1 (low load) | | | | | | |
| 50 | 60 | 64 | 65 | 93 | 58 | |
| 60 | 95 | 70 | 70 | 42 | 40 | |
| 40 | 65 | 70 | 30 | 60 | 92 | 2300 |
| 50 | 30 | 20 | 86 | 60 | 94 | |
| 90 | 76 | 60 | 50 | 70 | 80 | |
| 46 | 95 | 70 | 44 | 70 | 85 | |
| Scenario 2 (medium load) | | | | | | |
| 110 | 120 | 84 | 80 | 105 | 65 | |
| 70 | 105 | 80 | 75 | 100 | 98 | |
| 78 | 75 | 125 | 90 | 120 | 98 | 3300 |
| 60 | 90 | 80 | 95 | 115 | 100 | |
| 100 | 86 | 108 | 60 | 102 | 94 | |
| 66 | 105 | 94 | 104 | 78 | 85 | |
| Scenario 3 (high load) | | | | | | |

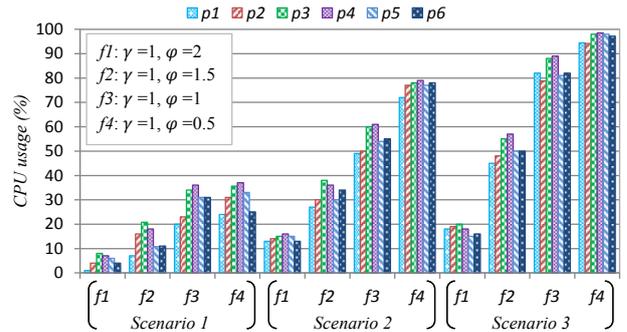

Fig. 8. The optimal CPU usages ($p_l$)

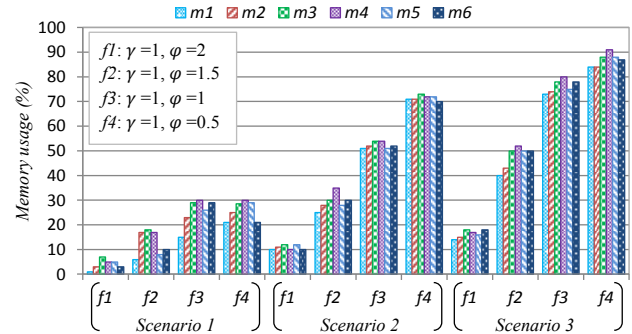

Fig. 9. The optimal memory usage ($m_l$)

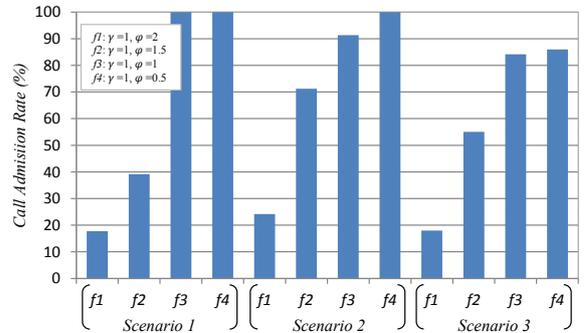

Fig. 10. The optimal call admission rate

For instance, in this scenario, in Case *f3*, the optimal value of $m_1$ and $p_1$ for admitting all input calls are 15.3214 and 20.64201, respectively. In Scenario 2, the offered load is higher than that of Scenario 1; the maximum call admission rate can be obtained in *f4* (Fig. 10). However, in this case, more resources are used compared to the previous scenario (Figs. 8 and 9). The comparison of Figs. 8, 9, and 10 indicates that in Scenario 3 even using the entire servers' resources, it is not possible to reach optimum call admission rate greater than 86%, as the input load would exceed the network capacity and the extra load would be blocked. By solving the model in Eq. (2), in addition to determining the optimal values of $m_l$, $p_l$, and $C^{ij}$, the optimal $R_{kl}^{ij}$ is also determined. For instance, in Scenario 3 and Case *f4*, the total call requests from path 1 to 6 ($\mathbb{C}^{16}$) is 65 in which 27.5 is admitted ($C^{16}$). Distribution of these admitted calls is shown in Fig. 11, $C^{16}$ is distributed between two paths to reach the maximum objective function as path 1: ($R_{13}^{16} = 11.2, R_{35}^{16} = 11.2, R_{56}^{16} = 11.2$) and path 2: ($R_{12}^{16} = 16.3, R_{24}^{16} = 16.3, R_{46}^{16} = 16.3$). For all Cases *f1* to *f4*, the load is distributed among the servers in a way that the maximum value for the objective function is achieved. The average time for each run ($t_c$) is almost 0.95 seconds, which can be ignored compared to the call length.

### B. Implementation and the Experimental Results

This section is organized in 8 subsections. We describe our implementation details in Section V.B.1. In Section V.B.2 measurements of $\alpha$ and $\beta$ are mentioned. In Section V.B.3 to V.B.5, we evaluate the VLB-CAC and give comprehensive results. Performance evaluation of VLB-CAC with autoscaling (Section V.B.6 and V.B.7), and the effect of time slot $\tau$ (Section V.B.8) are documented in the remaining subsections.

*1) Practical considerations and configurations*

To evaluate the performance of the proposed VLB-CAC in a real environment, we implanted it in the Natural Sciences and Engineering Research Council of Canada Strategic Network for Smart Applications on Virtual Infrastructure (SAVI) [11]. The SAVI project outlines a cloud system composed of two cloud types: core and smart edge (Fig. 12). As shown in Fig. 13, SAVI nodes incorporate open source software and hardware including OpenStack, OpenFlow and NetFPGAs [11]. Janus is the super controller of SAVI testbed. It controls converged heterogeneous resources (OpenStack, OpenFlow, FPGAs, GPUs). While Janus is controlling SAVI testbed, it utilizes Whale (a topology manager). Whale collects cloud computing resource information using OpenStack and networking resource information using OpenFlow and provides the status of all resources and their connectivity to Janus.

We define two different topologies namely Local and Wide shown in Figs. 14 (A) and (B), respectively. In the Local topology, all VMs are located at the University of Toronto (UofT) edge cloud while in the Wide topology, each VM is located on a different edge of the SAVI testbed.

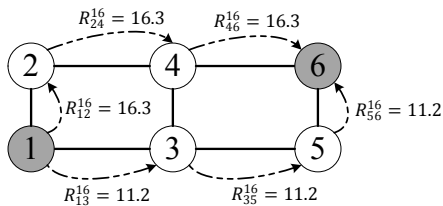

Fig. 11. Load distribution in paths among servers 1 through 6 in Scenario 3 Case *f4*, (signaling paths)

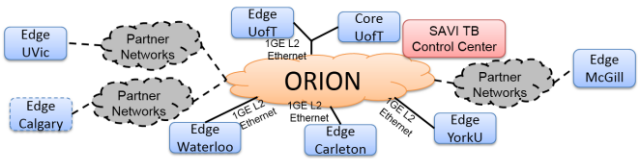

Fig. 12. Deployment of SAVI testbed in Canadian universities [11]

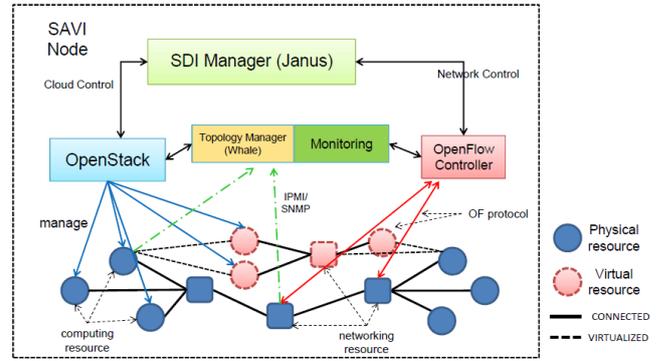

Fig. 13. Structure of a SAVI node [11]

We use the open source Elastix v.3 software [30] to implement the SIP servers on VM1 to VM6. On VM7, the open source SIPp software [31] is used as well to implement the user agents and inject the traffic ($\mathbb{C}^{ij}$). We implemented the proposed VLB-CAC on VM8, using PuLP [32]. This software is used to solve the mathematical models. PuLP is an LP modeler written in Python. We used Python, version 3.4. Furthermore, to make a connection between SIP servers, SIP trunk was used. For example, in VM1, only two SIP trunk connections to VM2 and VM3 have been defined. We use Centos as the operating system of VMs 1 to 6 and Ubuntu for VMs 7 and 8.

Each VM in the SAVI testbed can have one of the four flavors shown in Table III. All the VMs 1-8 have homogeneous features and their initial flavor is m1.small. By running the following command, each VM can be resized: `nova resize <VM Instance Name> <new flavor>`. For example, by running the command `nova resize VM1 m1.medium`, the VM 1 flavor can be changed to m1.medium.

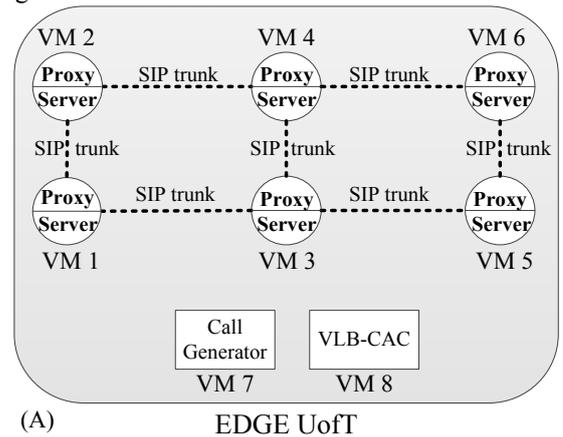

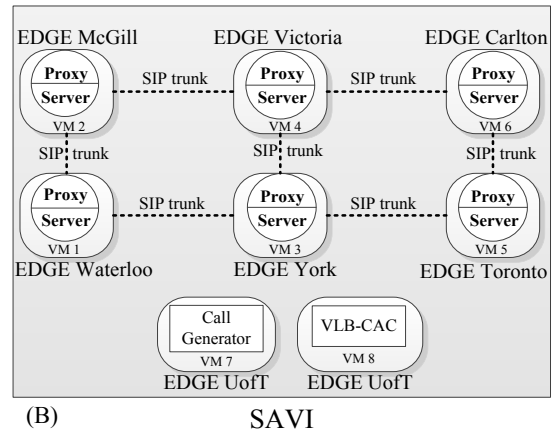

Fig. 14. (A): Local topology, (B): Wide topology

TABLE III. AVAILABLE VM FLAVORS SPECIFICATIONS ON SAVI TB

| Flavor | Memory (MB) | vCPUs | Disk (GB) |
|---|---|---|---|
| m1.small | 2048 | 1 | 20 |
| m1.medium | 4096 | 2 | 40 |
| m1.large | 8192 | 4 | 80 |
| m1.xlarge | 16384 | 8 | 160 |

The reports of Elastix v.3 software are used to measure call status, and OProfile software [33] is utilized to measure CPU and memory usages. The values of the CPU and memory usages ($P_i$ and $M_i$) are transferred to VLB-CAC through the use of Psutil tools [34] on each SIP server. Psutil (Python system and process utilities) is a cross-platform library for retrieving information on running processes and system utilization (CPU, memory, disks, and network) in Python.

Elastix v.3 software is a SIP server with an SIP proxy called SIP Proxy Kamailio (OpenSER) [35]. The proxy node is responsible for directing the call requests while the server node is responsible for responding to them. The optimal values of $C^{ij}$ and $R_{il}^{kj}$ calculated by VLB-CAC are transferred to VMs using UDP messages and are set in its kamailio configuration file (kamailio.cfg). It is important to note that if these values are non-integer, they will be rounded down and then sent to the nodes.

Suppose a call request ("INVITE" message) has reached an SIP server on VM $i$. Kamailio detects the message's initial source and final destination based on the message field headers:

▪ If both source and destination of the call are node $i$:

If $C^{ii} > 0$, this message is delivered to Elastix SIP server on VM $i$ to be handle and $C^{ii}$ is decreased by 1. If $C^{ii} = 0$, then this message is dropped in the fastest time possible before sending the "180 ringing" message. This makes the amount of resources that the SIP server spends for the dropped calls almost insignificant. This is also evident from the reported results.

▪ If the request source is node $i$ and the destination is node $j$:

If $C^{ij} > 0$, the value of $R_{il}^{ij}, l \in \{i\ neighbors\}$ is checked for all $l$s. If there are several $R_{il}^{ij} > 0$, one is chosen randomly and the "INVITE" is sent to node $l$ through the related SIP trunk. Then, $R_{il}^{ij}$ and $C^{ij}$ are decreased by 1. If $C^{ij} = 0$, then this message is dropped as suggested previously.

▪ If the request source is node $k$ and the destination is node $j$:

The value of $R_{il}^{kj}, l \in \{i\ neighbors\}$ is checked for all $l$s and if there are several $R_{il}^{kj} > 0$, one is chosen randomly and the "INVITE" message is sent to node $l$ through the related SIP trunk.

▪ If the request source is node $k$ and the destination is node $i$:
This message is delivered to Elastix node $i$ to be handled.

*2) Measurement of α and β*

To compute coefficients $\alpha$ and $\beta$, random inbound and outbound calls were established in the SIP server on VM 1 and the CPU and memory usages of the server were measured. This experiment was repeated for 100 times and the results were gathered in a dataset. By solving Eq. (3) and (4), $\alpha_1$ and $\alpha_2$ were determined as 0.07841 and 0.02158, and $\beta_1$ and $\beta_2$ were obtained 0.06998 and 0.01997, respectively. In the following, each experiment is conducted in a period of $\tau = 3s$ unless mentioned otherwise.

*3) Effect of VLB-CAC on the server's performance*

Fig. 15 shows the comparison of the SIP servers' performances in both local and Wide topologies in the presence and absence of VLB-CAC. Fig. 15 (A) and (B) (without VLB-CAC bar graph) confirm that if the incoming calls and network resources of the SIP servers are not monitored, the increase in offered load will cause resource saturation (Scenario 3). In this case, the servers are full of "INVITE" messages and send a retransmission request for the lost messages (Fig. 15 (C), without VLB-CAC bar graph). As already mentioned, this leads to more consumption of resources, increases the average call setup delay (the time between sending "INVITE" from the user agent and receiving "200 OK" from the SIP server) and causes a sharp drop in throughput (the number of serviced calls in a time unit) (Fig. 15 (D) , (E), without VLB-CAC bar graph). The drop in the number of serviced call steadily increases the retransmission rate (Fig. 15 (C), without VLB-CAC bar graph) and the resource consumption (Fig. 15 (A), (B), without VLB-CAC bar graph) and worsens the situation.

In the presence of VLB-CAC, servers will never face resources saturation (Fig. 15 (A), (B)) and, therefore, even in the case of overload (Scenario 3), server performance will not drop and the server will not suffer the consequences of the overload (Fig. 15 (E)). For example, in Case *f4* in Scenario 3 and in the Wide topology, without a sudden increase of the retransmission rate and the average call setup delay (Fig. 15 (C) , (D)), by spending an average of 98% of the CPU and 92% of the memory of all SIP servers, 2837 calls are admitted from 3300 call requests, and 2727 calls are serviced (Fig. 15 (E)), whereas in the absence of VLB-CAC in this scenario, only 750 call requests are serviced (Fig. 15 (E), without VLB-CAC bar graph).

Moreover, as shown in Table II, in Case *f2* of the second scenario, there are 2300 call requests created in total among which 1683 are admitted by VLB-CAC, and thus the admission rate reaches to 71.22% (Fig. 10). In the Local topology, among 1638 calls, only 49 calls do not end successfully while 1589 calls are serviced successfully (Fig. 15 (E)). Also in the Wide topology, from among 1638 calls, only 88 calls did not end successfully while 1550 calls are serviced successfully. According to Fig. 15 (E) repeating this experiment will clearly show that rounding the VLB-CAC values does not affect the number of the successfully serviced calls since the constraints of the proposed model in Eq. (2) are still in a feasible region.

Therefore, in the absence of VLB-CAC, since the resources are fully occupied, throughput is very low. While in the presence of VLB-CAC, an optimum throughput can be achieved through resource management. The results given in Fig. 15 (D) show that in Scenario 1, the average call setup delay is a little more in the presence of VLB-CAC. This is due to the time spots of $t_g$, $t_c$, and $t_n$ in VLB-CAC. VLB-CAC is able to prevent the retransmission rate increase and control the average delay in Scenarios 2 and 3 through spending these time spots.

*4) Simulation & implementation results comparison*

In this subsection, we compare the simulation and implementation results together to evaluate the validation of both the simulation and the implementation models. Figs. 16 and 17, show resource consumption of all VMs for Cases *f2* and *f4* in Scenario 2 (medium load). The results reported in these figures confirm that the performance of the SIP servers in both Wide and Local topologies are very close to that obtained by the simulation. Note that as SAVI uses the high-speed Layer 2 switching with 10GB links in its testbed, the end to end delay and the processing time is negligible, so the results for both Local and Wide topology are close to each other.

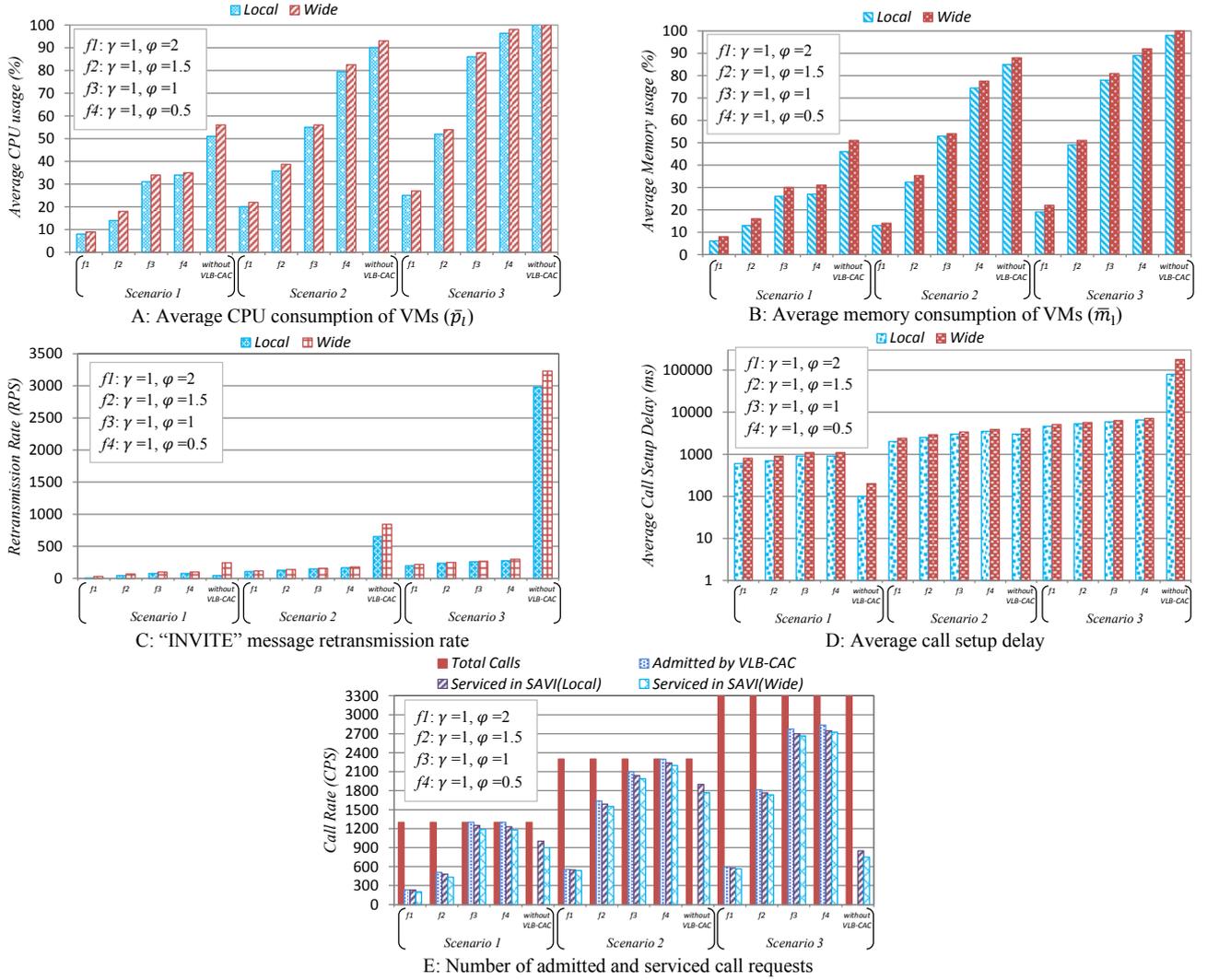

Fig. 15. Comparison of the results obtained in SAVI testbed with and without VLB-CAC

*5) Resizing the virtual machines*

As shown in Fig. 15 (E), VLB-CAC reaches an admission rate of 100% in both Scenarios 1 and 2, but as already noted in Fig. 10 and Fig. 15 (E), in Scenario 3, an optimum admission rate greater than 86% cannot be achieved due to the network resources input overload. In this situation, to achieve a greater admission rate and consequently greater service rate, we can use the resizing capability of the VM resources in the SAVI testbed. By changing all VMs from small to medium flavor and for Wide topology, Scenario 3 was performed with $\tau = 3s$. For this flavor, the new values of $\alpha_1$ and $\alpha_2$ are 0.08012 and 0.02329 and for $\beta_1$ and $\beta_2$ are 0.07169 and 0.02168, respectively. In Fig. 18, before and after the flavor change, the CPU and memory usages and the number of admitted calls are shown.

After resizing, VLB-CAC can achieve a call admission rate of 100% for Cases *f3* and *f4* of Scenario 3, (Fig. 18 (E)). As shown in Fig. 18 (F) for Cases *f3* and *f4* in Scenario 1 and Case *f4* in Scenario 2, the call rejection rate is zero. In Scenario 3, by resizing the VMs, the call rejection rate can reach zero (Fig. 18 (F), Scenario 3 Cases *f3* and *f4*). Furthermore, the comparison of the results obtained before and after resizing confirm that more call requests can be admitted and serviced after resizing.

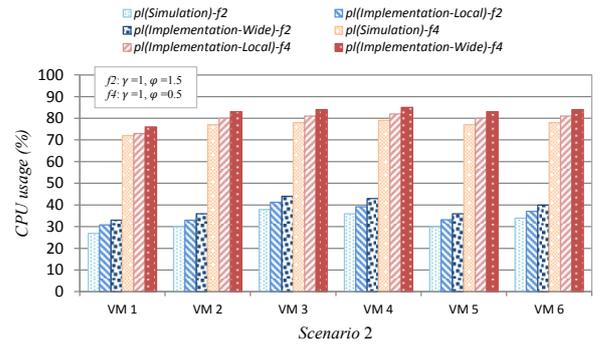

Fig. 16. Comparison of CPU usage in both simulation and implementation mode for Local and Wide topologies for two Cases *f2* and *f4*

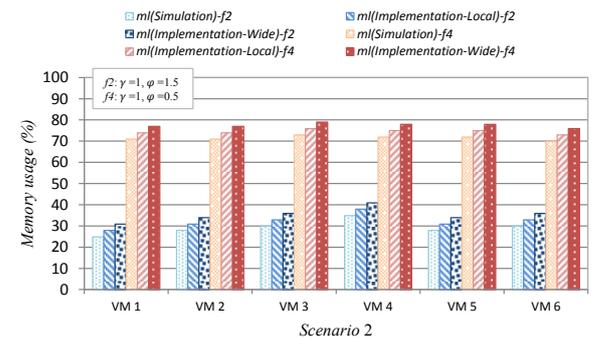

Fig. 17. Comparison of memory usage in both simulation and implementation mode for Local and Wide topologies for two Cases *f2* and *f4*

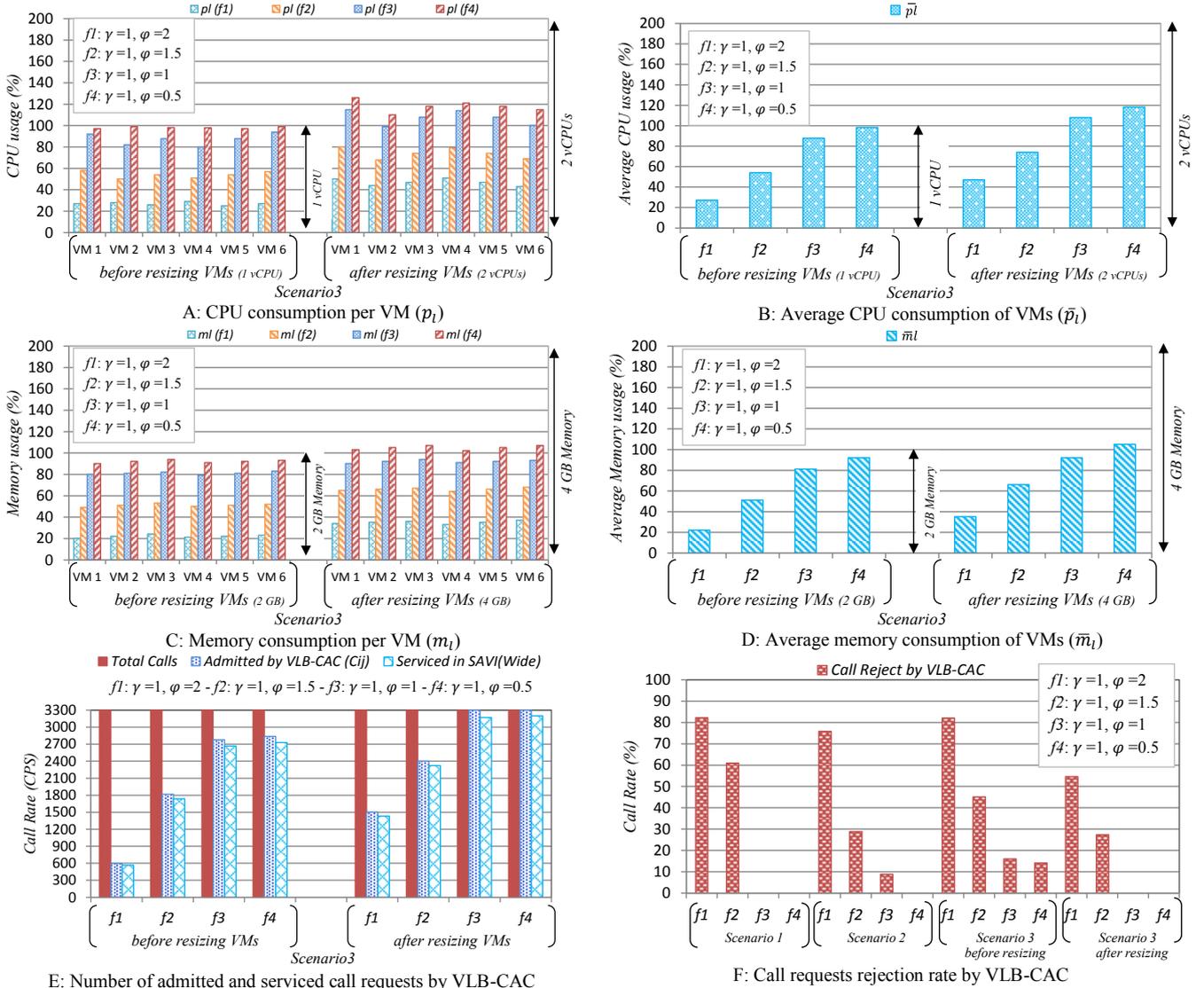
Fig. 18. Results of resizing VMs in SAVI testbed with VLB-CAC (Wide topology)

For example, in Case *f1* of Scenario 3, the values of $\bar{p}_l$ before and after resizing are equal to 27% and 47% (Fig. 18 (B)), respectively; the values of $\bar{m}_l$ before and after resizing are equal to 22% and 35% (Fig. 18 (D)), respectively. The number of admitted calls before and after resizing are equal to 594 and 1500, respectively. The number of serviced requests in the Wide topology before and after resizing are equal to 564 and 1430 (Fig. 18 (E)), respectively. Therefore, resources can be virtualized to increase throughput in SIP networks.

In Fig. 18, parts (A) to (D), it is emphasized that none of $p_l$s and $m_l$s have a significant difference with $\bar{p}_l$s and $\bar{m}_l$s. It means that the traffic has been uniformly distributed between all servers. Note that the mean and standard deviation of $\mathbb{C}^{ij}$ in Scenario 3 are equal to 91.66 and 14.24, respectively. Furthermore, all VMs have the same flavor before and after resizing. An example of a load distribution is provided in Fig. 19 (A). The distribution of $\mathbb{C}^{16}$ before resizing in the Wide topology in Case *f4* of Scenario 3 is shown in this figure. The simulation results of this case were already shown in Fig. 11.

A comparison between Figs. 19 (A) and 11 will lead to this understanding that the implementation results are very close to the simulation results and therefore, the settings of the configuration files (kamailio.cfg) of each node proxy for directing the admitted calls has been done properly. Among 65 requests ($\mathbb{C}^{16}$), 27.5 ($C^{16}$) requests are admitted and routed by VLB-CAC (Fig. 11), but before sending the results to the servers, the values of $C^{ij}$ and $R_{kl}^{ij}$ are rounded down. At the end of $\tau$, a total of 25 "INVITE" requests are serviced and distributed between two paths (Fig. 19 (A)). After resizing, all the 65 requests are admitted and 62 requests are serviced based on the distribution shown in Fig. 19 (B).

*6) Autoscaling of virtual machines*

All previous experiments were only conducted in a period of $\tau = 3$s. This section examines the carried load and the average resource consumption in Case *f4* in the Wide topology. We evaluate the system for a time interval equal to 3000 seconds. Two different cases $\tau = 3s$ and $\tau = 3.5s$ are considered (Fig. 20). In case $\tau = 3s$ and $\tau = 3.5s$, we have 1000 and 857 periods, respectively. The initial flavor of all VMs is set to small. To evaluate the performance of the system under the sudden change in the offered load, five different scenarios were run tandem, each for 600s. As it can be seen in Fig. 20 (A), at the first 600s, a low load (Scenario 1) is injected; at t = 600s, the offered load is increased by applying Scenario 2 with the medium load. At t = 1200s, a high load (Scenario 3) is injected to the system. After that at the next two 600s time intervals, the offered load is reduced to Scenarios 2 and 1, respectively.

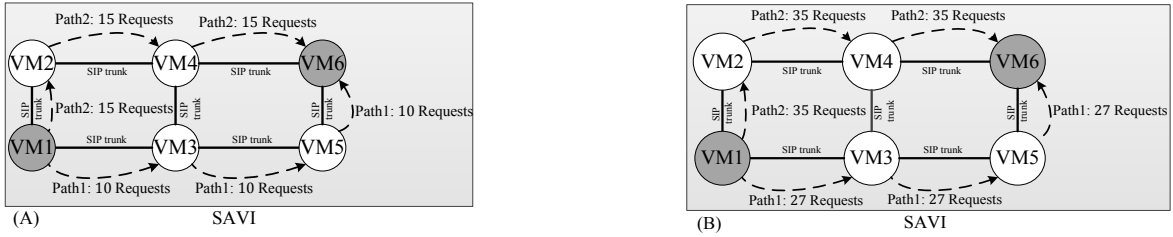

Fig. 19. Load distribution in paths among servers 1 through 6 (Wide topology), (A): Before resizing VMs, (B): After resizing VMs

It can be seen that in all scenarios, by using VLB-CAC, we can achieve a high carried load close to the offered load (Fig. 20 (A)), while we consume fewer resources compared to the case when we do not use VLB-CAC ("Without VLB-CAC" curve in Fig. 20 (B), (C)). Furthermore, in all scenarios with VLB-CAC, the carried load in case $\tau = 3s$ is a little bit more than the case $\tau = 3.5s$ since in the case $\tau = 3s$, the hold-on time for the new calls is less and the intervals of VLB-CAC decisions is shorter and consequently $R_{kl}^{ij}$ and $C^{ij}$ are more accurate. However, as shown in Fig. 20 (B), (C), the CPU and memory consumption in case $\tau = 3s$ is more than that of case $\tau = 3.5s$.

Moving from Scenario 1 to Scenario 2, the offered load is increased and the carried load in both cases is also increased. At $t = 1200s$, a flash crowd occurs and the offered load changes to the high load scenario continue for 600 seconds. Flash crowd occurs when a large number of users simultaneously attempt to make a call. Several factors affect the occurrence of the flash crowd. For example, on television voting or on special days like New Year's Day, a large number of calls are placed at short intervals imposing an overload on the network. In the case of "Without VLB-CAC", despite the full occupation of the resources (Fig. 20 (B), (C)), the carried load is dropped sharply and the network suffered the consequences of the overload (Fig. 20 (A)). The overload is so severe that by reducing the offered load at $t = 1800s$, the network service rate does not return to its normal case. This is because the resources are not completely released due to their participation in message retransmissions (Fig. 20 (B), (C), the fourth 600 seconds).

As discussed in Section V.B.5, in the case of "With VLB-CAC" in Scenario 3, since the input load is higher than the network resources, a throughput very close to the offered load cannot be achieved (Fig. 20 (A)) even by spending more resources (Fig. 20 (B), (C)). This issue can be resolved by increasing the server resources.

We have equipped VLB-CAC with proposed autoscaling scheme in Section IV.B. Autoscaling enables VLB-CAC to automatically scale up the SIP server when facing overload condition and scale down when the condition is normal. At the start of Scenario 3 (t = 1200s) and in the case of "With VLB-CAC ($\tau = 3s$)", call admission rate, $\bar{p}_l$ and $\bar{m}_l$ are respectively 86%, 98% (of 1 vCPU) and 92% (of 2 GB memory).

In the case of "With VLB-CAC and Autoscaling ($\tau = 3s$)", the VMs flavor is upgraded from small to medium (2 vCPUs and 4 GB memory). In this way, call admission rate, $\bar{p}_l$ and $\bar{m}_l$ for the new flavor are 100%, 62% (of 2 vCPUs) and 55% (of 4 GB memory), respectively. As a result, among 3300 call requests, a total number of 3190 (Fig. 20 (A), ($\tau = 3s$)) or 3050 (Fig. 20 (A), ($\tau = 3.5s$)) calls are serviced from t = 1200s to t = 1800s. As at t = 1800s, the offered load drops, the VMs flavor is changed from medium to small and the additional resources are released. In the last two scenarios (from second 1800 to 3000), the carried load can be raised nearly to the offered load without changing the flavor.

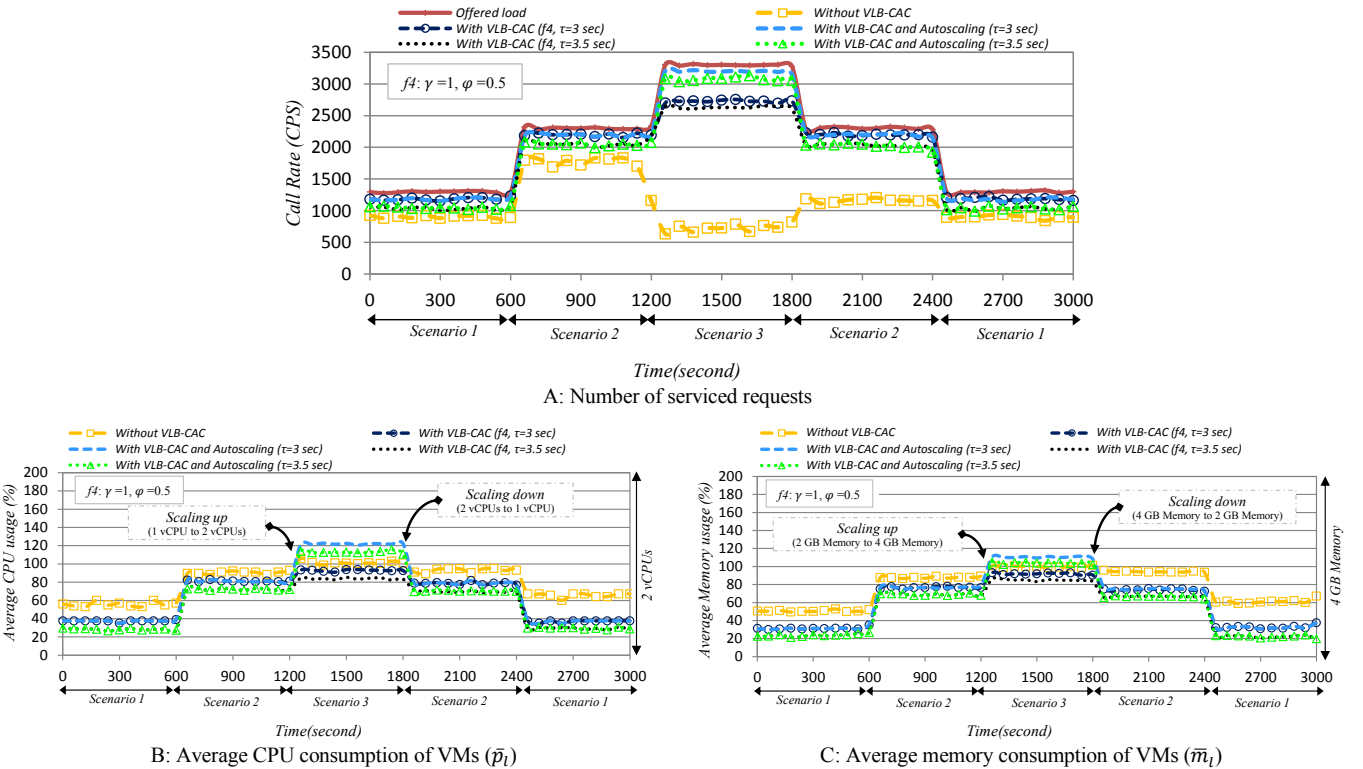

A: Number of serviced requests

B: Average CPU consumption of VMs ($\bar{p}_l$)

C: Average memory consumption of VMs ($\bar{m}_l$)

Fig. 20. SIP Network performance over time as the offered varies (Wide topology)

*7) Effect of node failure*

In addition to the flash crowd, another reason causing the overload of the SIP servers is the sudden failure of the network nodes and the reduction of network capacity. This failure may be due to the load imposed by the failed server on the others in the network. To evaluate the performance of VLB-CAC when facing this situation, we inject $\mathbb{C}^{16} = 1000$ calls (offered load) to the Wide topology over a period of 600 seconds. At $t = 200s$, VM3 fails and at $t = 400s$, it goes back to service again. The results are shown in Fig. 21.

As can be seen in Fig. 21 (A), by $t= 200s$, an average of 955 requests is serviced with VLB-CAC. When VM3 is shutdown at $t = 200s$, there is only one available path for $\mathbb{C}^{16}$ (path 2). In this condition, VLB-CAC intends to use most of the server capacity in the existing path without disturbing consequences of the overload to cope with the sharp drop of the service rate. That is why $\bar{p}_l$ and $\bar{m}_l$ increase approximately to 97% and 92% (Fig. 21 (B), (C)). However, at $t = 200s$, the carried load drops to 637 calls and the average call admission rate reaches 64.5% (Fig. 21 (A)). The use of autoscaler in VLB-CAC provides the requirements for resizing the VMs in path 2 from small to medium.

In this way, at $t = 200s$, the carried load reaches to 955 without any drop. When VM3 returns to the service at $t = 400s$, the overload situation is passed. The call admission rate, $\bar{p}_l$ and $\bar{m}_l$ for medium flavor reaches approximately 100%, 31% (of 2 vCPUs) and 29% (of 4 GB memory). That is why the VMs flavor change from medium to small. Thereafter, all the VMs work with small flavor and can return to the service rate before the overload (Fig. 21). However, as already mentioned, in case "Without VLB-CAC", there is no way to reach the service rate before the overload even after passing the overload.

*8) Effect of duty cycle $\tau$*

In this subsection, we investigate the effect of VLB-CAC's duty cycle ($\tau$) on the system performance. For this purpose, we evaluate the call admission rate and resource utilization of the VLB-CAC in different values of $\tau$ ($\tau = 2, 4, 6, 8s$). As shown in Table IV, when $\tau$ is low ($\tau = 2s$) the hold-on time for the new calls is low and the interval of VLB-CAC decision is short and consequently, $R_{kl}^{ij}$ and $C^{ij}$ are more accurate which causes a higher call admission rate. However, as shown in Table IV, the average CPU and memory consumption in case $\tau = 2s$ are more than those of the others cases.

TABLE IV. COMPARISON OF THE RESULTS AS THE $\tau$ IS VARIED WITH VLB-CAC (SCENARIO 1 OF WIDE TOPOLOGY)

| | f2 | | | | f4 | | | |
|---|---|---|---|---|---|---|---|---|
| $\tau$ (s) | 2 | 4 | 6 | 8 | 2 | 4 | 6 | 8 |
| Call admission rate (%) ~ | 41 | 37 | 21 | 14 | 100 | 96 | 80 | 74 |
| Average CPU usage (%) ~ | 22 | 20 | 17 | 13 | 38 | 35 | 32 | 27 |
| Average memory usage (%) ~ | 19 | 17 | 15 | 13 | 31 | 28 | 24 | 20 |

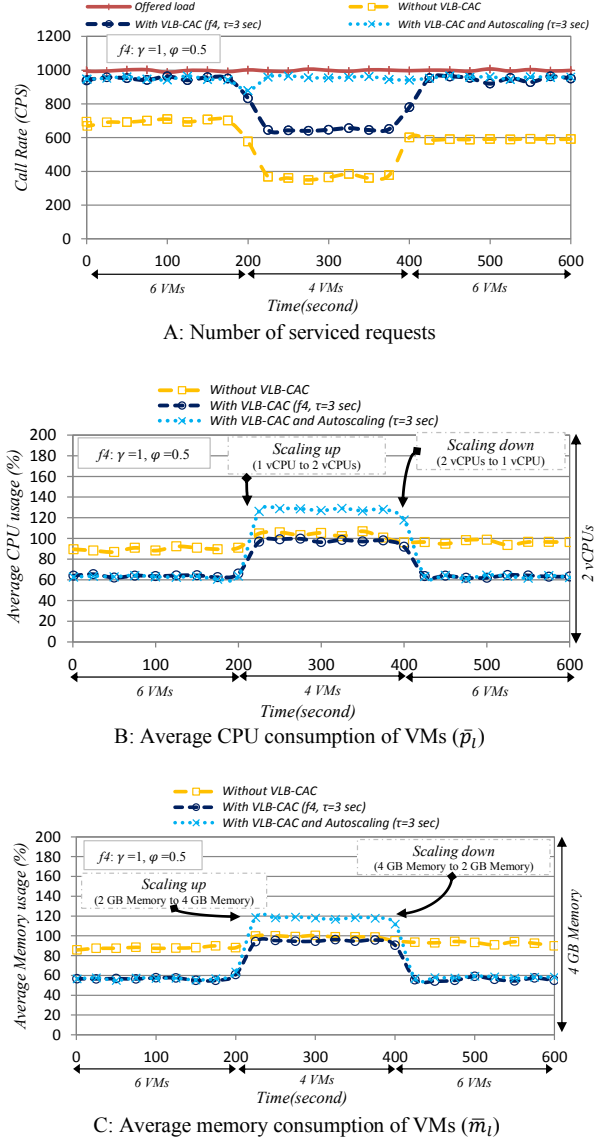

A: Number of serviced requests

B: Average CPU consumption of VMs ($\bar{p}_l$)

C: Average memory consumption of VMs ($\bar{m}_l$)

Fig. 21. Network performance over time in case of VM3 failure

### C. Comparison with Other Algorithms

In this section, we assess the performance of the algorithm implemented on VLB-CAC (Section IV.A) against [22] and [4] (the configuration is similar to that of Section V.A). In [4], TLWL algorithm routes a new call request to the server with the least load using a load balancer. On the other hand, [22] employs a mixed-integer nonlinear optimization on three levels to allocate resources with respect to the cloud platform input requests (here, we consider medium level which is referred to as $RAP_c^{T_2}$). "Algorithm execution time", "resource consumption for running the algorithm as the cost of the algorithm" and "throughput as the algorithm output" are considered for the assessment. As Table V shows, VLB-CAC is able to achieve a better throughput compared with the other two algorithms, which happens as a result of the linearity of the model. Moreover, linearity allows lower execution time and resource consumption.

TABLE V. VLB-CAC PERFORMANCE COMPARISON WITH OTHER ALGORITHMS

| The number of servers and incoming requests | $n = 6$ and $\sum_{i=1}^{n}\sum_{j=1}^{n}\mathbb{C}^{ij} = 3{,}300$ | | | $n = 12$ and $\sum_{i=1}^{n}\sum_{j=1}^{n}\mathbb{C}^{ij} = 6{,}600$ | | | $n = 24$ and $\sum_{i=1}^{n}\sum_{j=1}^{n}\mathbb{C}^{ij} = 13{,}200$ | | |
|---|---|---|---|---|---|---|---|---|---|
| Algorithms | VLB-CAC | $RAP_c^{T_2}$ [22] | TLWL [4] | VLB-CAC | $RAP_c^{T_2}$ [22] | TLWL [4] | VLB-CAC | $RAP_c^{T_2}$ [22] | TLWL [4] |
| Algorithm Execution Time (s) ~ | 0.42 | 0.49 | 3.24 | 0.48 | 4.89 | 32.68 | 0.54 | 7.15 | 120.7 |
| CPU Consumption (%) ~ | 3 | 8 | 14 | 4 | 15 | 19 | 5 | 26 | 38 |
| Memory Consumption (%) ~ | 1 | 7 | 10 | 2 | 13 | 15 | 3 | 24 | 28 |
| Throughput (req/s) ~ | 2,838 | 2,475 | 2,244 | 5,676 | 4,950 | 4,488 | 11,352 | 10,032 | 8,976 |

As the number of servers and requests increase, VLB-CAC is able to demonstrate better scalability. By increasing servers, the number of variables in $RAP_c^{T2}$ increase rapidly, which in turn demands more memory and processor. It should also be noted that LP can be solved with great efficiency in polynomial time by interior point methods, yet MINLPs are NP-hard (see [25, 26]).

To compare our algorithm with TLWL, note that TLWL requires preserving the information from the received request for load estimation (and the whole signaling traffic of SIP network passes through the load balancer). Hence, the required resources for TLWL would burgeon by increasing the number of requests. As mentioned in [4], TLWL offers scalable solution for a maximum of 10 servers.

## VI. Conclusion and Future Work

In this paper, we proved that the problem of overload control in SIP network with $n > 2$ servers and in limited resources is NP-hard. We introduced a Virtual Load-Balanced Call Admission Controller (VLB-CAC) based on a heuristic mathematical model to determine optimal resource allocation in such a way that the number of requested local and outbound calls are maximized regarding the limited resources of the SIP servers. Specifically, we proposed a linear optimization model to maximize call admission rates along with the optimal allocation of CPU and memory resources of the SIP servers. The under-study SIP network was implemented in the VMs through the use of virtualization capacity in the SAVI testbed and VLB-CAC was equipped with an autoscaling method to overcome the resource limitations. We developed the various autoscaling policies to deal with the SIP network overload. An assessment of the analytical and experimental results in various scenarios demonstrates the efficiency of the proposed method. As future work, we intend to make effective use of SDNs (Software Defined Networks) for controlling overload in SIP networks. We also plan to implement VLB-CAC using the OpenFlow protocol.

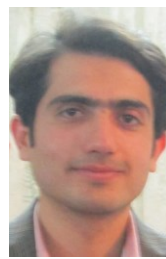

**Ahmadreza Montazerolghaem** received the B.Sc. degree in Information Technology from the computer department, Sadjad University of Technology and M.Sc. degree in computer engineering from the computer department, Ferdowsi University of Mashhad (FUM), Iran, in 2010 and

2013, respectively. Currently, he is a Ph.D. candidate in computer engineering at computer department, FUM. He is an IEEE Student member and a member of IP-PBX type approval lab in FUM. He is also a member of National Elites Foundation (Society of prominent students of the country). His research interests are in Software Defined Networking, Network Function Virtualization, Voice over IP, and Optimization.

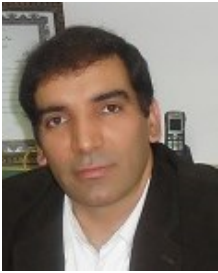

**Mohammad Hossein Yaghmaee Moghaddam** received his B.S. degree in communication engineering from Sharif University of Technology, Tehran, Iran in 1993, and M.S. degree in communication engineering from Tehran Polytechnic (Amirkabir) University of Technology in 1995. He received his Ph.D degree in communication engineering from Tehran Polytechnic (Amirkabir) University of Technology in 2000. He has been a computer network engineer with several networking projects in Iran Telecommunication Research Center (ITRC) since 1992. November 1998 to July1999, he was with Network Technology Group (NTG), C&C Media research labs., NEC corporation, Tokyo, Japan, as visiting research scholar. September 2007 to August 2008, he was with the Lane Department of Computer Science and Electrical Engineering, West Virginia University, Morgantown, USA as the visiting associate professor. July 2015 to September 2016, he was with the electrical and computer engineering department of the University of Toronto (UoT) as the visiting professor. Currently, he is a full professor at the Computer Engineering Department, Ferdowsi University of Mashhad (FUM). His research interests are in Smart Grid, Computer and Communication Networks, Quality of Services (QoS), Software Defined Networking (SDN) and Network Function Virtualization (NFV). He is an IEEE Senior member and head of the IP-PBX type approval lab in the Ferdowsi University of Mashhad. He is the author of some books on Smart Grid, TCP/IP and Smart City in Persian language.

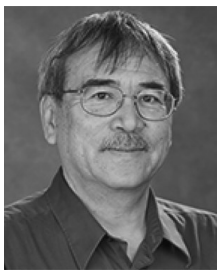

**Alberto Leon-Garcia** received the B.S., M.S., and Ph.D. degrees in electrical engineering from the University of Southern California, Los Angeles, CA, USA, in 1973, 1974, and 1976, respectively. He was founder and CTO of AcceLight Networks, Ottawa, ON, Canada, from 1999 to 2002, which developed an all-optical fabric multiterabit, switch. He is currently a Professor in Electrical and Computer Engineering at the University of Toronto, ON, Canada. He holds a Canada Research Chair in Autonomic Service Architecture. He holds several patents and has published extensively in the areas of switch architecture and traffic management. His research team is currently developing a network testbed that will enable at-scale experimentation in new network protocols and distributed applications. He is recognized as an innovator in networking education. In 1986, led the development of the University of Toronto-Northern Telecom Network Engineering Program. He has also led in 1997 the development of the Master of Engineering in Telecommunications program, and the communications and networking options in the undergraduate computer engineering program. He is the author of the leading textbooks Probability and Random Processes for Electrical Engineering and Communication Networks: Fundamental Concepts and Key Architecture. His current research interests include application- oriented networking and autonomic resources management with a focus on enabling pervasive smart infrastructure. Prof. Leon-Garcia is a Fellow of the Engineering Institute of Canada. He received the 2006 Thomas Eadie Medal from the Royal Society of Canada and the 2010 IEEE Canada A. G. L. McNaughton Gold Medal for his contributions to the area of communications.

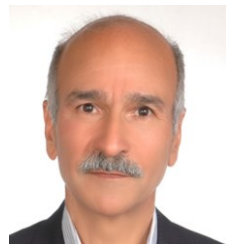

**Mahmoud Naghibzadeh** has received the MS and PhD degrees in Computer Science and Computer Engineering, respectively, from University of Southern California, USA. Now, he is a full professor at the Deparment of Computer Engineering, Ferdowsi University of Mashhad, Mashhad, Iran. In 1991 he was a visitng professor at University of California, Irvine, USA, and in 3003-2004 he was a visitng professor at Monash University, Australia. He is the director of Knowledge Engineering Reasearch Group (KERG) laboratory and his research interests include the schedulig aspects of real-time systems, Grid, Cloud, Multiprocessors, Multicores, and GPGPUs. Besides, he is also interested in Bioinfomatics computer algorithms, especially protein structures and protein-preotein interactions. He has published numerous papers in international journals and conference proceedings as well as eight books in the field of Computer Science and Engineering. He has been the general chair of two international computer conferences and the technical chair of many others. Also, he is the reviewer of many journals and member of many computer societies, especially a senior member of the IEEE. He is the recipiant of many awards including MS and PhD study scholarship and outstanding professor award. Currently, he is the chief editor of the Computer and Knowledge Engineering (CKE) journal.

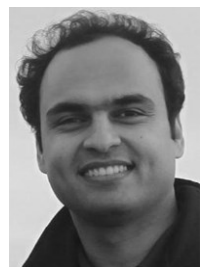

**Farzad Tashtarian** received the B.S. degree in Computer Engineering from Islamic Azad University Mashhad Branch, Iran in 2005, and the M.S. degree in Information Technology from Islamic Azad University Qazvin Branch, Iran in 2007. He received his Ph.D. degree in Computer Engineering at Ferdowsi university of Mashhad, Mashhad, Iran in 2013. Currently, he is an assistant professor and the Head of IT Department at IAUM. His research interests include wireless sensor networks, mobile communications, mathematical modeling, optimization, and distributed control.